\documentclass[journal, 12pt, final, onecolumn]{IEEEtran}

\usepackage{versions}
\includeversion{ARXIV}
\excludeversion{JSAC}

\usepackage[cmex10]{amsmath} \usepackage{amssymb}
\usepackage{enumerate} \usepackage{amsthm} \usepackage{pifont}
\usepackage{graphicx}

\newcommand{\cross}{\ding{55}}
\newcommand{\tick}{\ding{51}}

\theoremstyle{plain}
\newtheorem{mylemma}{Lemma}

\newtheorem{myexample}{Example}
\theoremstyle{theorem} \newtheorem{mythm}{Theorem}
\theoremstyle{definition} \newtheorem{mydef}{Definition}

\title{A Strategy-Proof and Non-monetary Admission Control Mechanism
  for Wireless Access Networks}

\author{\IEEEauthorblockN{Xiaohan Kang, Juan Jos\'e Jaramillo, Lei
    Ying}\\
  \IEEEauthorblockA{Department of Electrical and Computer Engineering\\
    Iowa State University, Ames, IA, 50011\\
    Email: \{xkang,jjjarami,leiying\}@iastate.edu}
      \thanks{This paper is an extended version of an earlier paper that appeared in \cite{Kang10}.}
      \thanks{Research supported by NSF Grant CNS-0953165, and DTRA Grants
HDTRA1-08-1-0016 and HDTRA1-09-1-0055.}
    }

\begin{document}
\maketitle
\begin{abstract}
  We study admission control mechanisms for wireless access networks
  where (i) each user has a minimum service requirement, (ii) the
  capacity of the access network is limited, (iii) the access
  point is not allowed to use monetary mechanisms to guarantee that
  users do not lie when disclosing their minimum service
  requirements, and (iv) the access point wants to admit as many users as
  possible. To guarantee truthfulness, we use auction theory to
  design a mechanism where users compete to be admitted into the
  network. We propose admission control mechanisms under which the
  access point intelligently allocates resources based on the
  announced minimum service requirements to ensure that users have no
  incentive to lie and the capacity constraint is fulfilled. We also
  prove some properties that any feasible mechanism should have.
\end{abstract}
\section{Introduction}
\label{introduction}
Resource allocation has been one of the most important issues in the
design of communication networks.  Given a wireless access network, in
which the users have various quality of service (QoS) requirements and the access point has
limited resources, admission control mechanisms are vital to achieve
stability, fairness and efficiency of the system. In this paper, we
consider a wireless access network with multiple users and a single
access point. We assume that the access point network is a public
network, and is not allowed to charge users for accessing the
network. We study the case when the QoS requirements are private, and
users are allowed to selfishly disclose any value that would give them
better service. Then the problem is to design an admission control
mechanism such that the true QoS of the users can be collected without
the use of any pricing scheme, and as many users as
possible are admitted. A natural choice is to set up a game with the users, such
that the selfish users are incentivized to tell the truth. Originated
in economics theory, auction mechanisms have been found very useful in
this kind of situations, since they are designed to entice selfish
bidders to tell the truth when allocating limited resources.

Various auction mechanisms have been well-studied. Myerson \cite{mye}
has obtained the optimal auction mechanism in closed form mathematical
expression. His result, however, only works for limited utility
functions without constraints over the resource. The well-known VCG
mechanism has been proved to guarantee truth-telling while achieving
the social optimum \cite{kri}. The challenge of designing an effective
auction mechanism in the proposed setting is the access point is not
allowed to charge the users, so the auction mechanism has to be
non-monetary. Thus, the VCG mechanism cannot be easily adapted in our scenario due to
the non-monetary requirement. Credit schemes that were developed to
incentivize cooperation in wireless networks could also be adapted to
guarantee users do not lie about their true requirements, but they
would also require secure mechanisms to avoid tampering with the
virtual money \cite{and:eid}, \cite{but:hub}, \cite{but:hub:2},
\cite{cgko}, \cite{sncr}, \cite{sncr:2}, \cite{zcy}. Recently, Hou and
Kumar have proposed a bidding game between users and access point that
maximizes the total utility, but in the iterative process the users
are forced to bid specific values instead of bidding selfishly
\cite{hou:kum}. In \cite{agg:har} a knapsack auction is studied, where
multiple bidders want to place objects of different sizes and valuations.
While the problem also has capacity constraints, the object sizes are public knowledge and the auction is allowed
to use payment schemes to guarantee users truthfully reveal their private valuations. A non-monetary
mechanism is studied in \cite{meo:mil}, where only the user that requests
the smaller service rate is admitted. This model is useful in peer-to-peer
networks if the access technology does not provide separation between
the upstream and downstream flows and users want to minimize the
upload bandwidth in order to increase their download bandwidth. However, in the access point model that we
study, the goal is to admit as many users as possible without violating
the capacity constraint. In this paper, we seek to design auction mechanisms
that are truthful and do not use any money-based scheme for general
utility functions.

Our contributions are therefore threefold.
\begin{enumerate}[i.]
\item We model the admission control problem, whose objective is to admit as many users as possible, as an auction mechanism
  design with resource constraint.
\item Second, we present two theorems that help us understand the
  essence of strategy-proof mechanisms. The first theorem
  (Theorem~\ref{impossibility}) shows the impossibility of a
  reasonable truth-telling mechanism to be based on probabilistic
  decisions, and the second theorem (Theorem~\ref{lwbthm}) shows that
  the truthfulness of a mechanism is equivalent to the existence of a
  highest winning bid regardless of one's own bid.
\item We propose our mechanism and show that it has the desired
  properties and admits at least half of the optimal number of users
  with high probability in an asymptotic sense.
\end{enumerate}

This paper is organized as follows. Section~\ref{model} gives the
model of this problem and the set of assumptions. Section~\ref{analysis}
analyzes the problem and characterizes the properties for feasible
auction mechanisms. Section~\ref{mechanism} gives our proposed
mechanism and shows feasibility and the performance
bound. Section~\ref{conclusion} concludes this paper.
\section{Model}
\label{model}
Consider a multiple access network with $n$ users and a single access
point (AP), where only one user can get a certain amount of resource
allocated by the AP at any given time slot. Let $\mathcal N = \{1, 2,
\cdots, n\}$ be the set of all users. The AP is assumed to have
resources with the total amount of $C > 0$. Each user is assumed to
have a quality of service (QoS) requirement $q_i \le C$, which
indicates the resource requested by the user $i$ and is only known to
the user $i$. We suppose that the range of $q_i$ is $T_i = [a_i,
b_i]\subset[0, C]$ and let $T = T_1\times T_2\times\dots\times T_n$
and $T_{-i} = T_1\times T_2\times\dots\times T_{i-1}\times
T_{i+1}\times\dots\times T_n$ where $\times$ denotes the Cartesian
product. We use bold letters like $\mathbf q, \mathbf t, \mathbf s\in
T$ and $\mathbf q_{-i}, \mathbf t_{-i}, \mathbf s_{-i}\in T_{-i}$ to
imply vectors rather than scalars.

By setting $q_i' = q_i/C$ and taking $q_i'$ as the requested resource,
the total amount of resources would then be 1. So without loss of
generality we set $C = 1$ and call $q_i$ the service rate of user $i$
for the rest of the paper.

Our objective is to admit as many users as possible. Thus, in order to decide which users to serve and the QoS the AP should
provide, the AP sets up an auction, in which each user, or bidder,
bids a requested service rate $t_i\in T_i$, which can be different
from the true service rate $q_i$, and the AP decides the set of users
that get admitted and assigns service rate $x_i$ to user $i$. Note
that the bid $\mathbf t$ is different from the bids in a traditional
auction in that the higher a user bids, the more resource he/she is
requesting, and thus the less likely he/she should be admitted. By
Myerson's revelation principle \cite{mye}, we only consider direct
revelation mechanisms, i.e., the mechanisms in which users submit bids
in the form of service rate values rather than in the form of any
other strategies. Then the mechanism can be described by the
\emph{outcome functions} $(p, x)$, where $p$ and $x$ are the set of
functions
\begin{align*}
  p(\mathbf t) & = (p_\phi(\mathbf t), \phi \subset\mathcal N)\\
  x(\mathbf t) & = (x_{i, \phi}(\mathbf t), \phi\subset\mathcal N)
\end{align*}
with $p_\phi\colon T\to\mathbb R$ being the probability that only the
subset of users $\phi$ of $\mathcal N$ get admitted and $x_{i,
  \phi}\colon T\to\mathbb R$ being the actually assigned service rate
for user $i$ if the subset $\phi$ is admitted. For convenience we set
$x_{i, \phi}(\mathbf t)$ to be the assigned service rate by mechanism
$(p, x)$ if $i\in\phi$ and $p_\phi(t) > 0$, and 0 otherwise, i.e., no
resource is assigned to user $i$, which is equivalent to not admitting
user $i$.

The utility function for user $i$, given the true service rate $q_i$
and the assigned service rate $x_i$, is $u_i(q_i, x_i)$. It is assumed
that $u_i(\cdot,\cdot)$ is non-negative, non-decreasing with $x_i$ when $x_i \ge
q_i$, and equals 0 when $x_i < q_i$. 
Given mechanism $(p, x)$, the expected utility for user $i$ with true
service rate $q_i$, bid service rate $t_i$ and others bidding $\mathbf
t_{-i}$ is
\begin{equation}
  \label{utility}
  U_i(p, x, q_i, t_i, \mathbf t_{-i}) = \sum_{\phi:i\in\phi}u_i(q_i, x_{i,
    \phi}(\mathbf t))p_\phi(\mathbf t)\ .
\end{equation}


We now present some definitions we will use throughout the paper.

\begin{mydef}[Incentive compatibility - IC]
  A mechanism $(p, x)$ is \emph{incentive compatible} or \emph{truthful} or \emph{strategy-proof} if for any
  utility function, any $i\in\mathcal N$, any $\mathbf q\in T$ and any
  $t_i\in T_i$, we have
  \begin{equation}
    \label{ic}
    U_i(p, x, q_i, q_i, \mathbf q_{-i}) \ge U_i(p, x, q_i, t_i,
    \mathbf q_{-i})
  \end{equation}
  where $U(\cdot)$ is defined in (\ref{utility}). That is, any
  possible true service rate vector $\mathbf q$ is a Nash equilibrium
  \cite{fud:tir}, in which no user has incentive to lie if all the
  other users bid their true service rates.
\end{mydef}

It must be noted that the definition of IC requires no user has incentive to
lie \emph{regardless} of the utility function they have, as long as the
assumptions on $u_i(q_i, x_i)$ hold. This concept of user utility is then fundamental to
prove the results on truthfulness. However, it must be emphasized that
our objective is to admit as many users as possible and not to maximize the total
network utility.

\begin{mydef}[Weak-incentive compatibility - weak-IC]
  A mechanism $(p, x)$ is \emph{weakly-incentive compatible} if for
  any utility function, for any $i\in\mathcal N$ and any $t_i\in T_i$,
  we have
  \begin{equation}
    \label{weakic}
    U_i(p, x, q_i, q_i, \mathbf q_{-i}) \ge U_i(p, x, q_i, t_i,
    \mathbf q_{-i}),\quad\forall\mathbf q\in T^*
  \end{equation}
  for some $T^*\subset T$ with $\mathcal L(T\backslash T^*) = 0$,
  where $\mathcal L(\cdot)$ is the Lebesgue measure.\footnote{Lebesgue
    measure is the standard way to measure the subsets of an
    $n$-dimensional Euclidean space, which coincides with the standard
    measure of $n$-dimensional volumes.}
\end{mydef}
\begin{mydef}[Feasibility]
  A mechanism $(p, x)$ is \emph{feasible} if it satisfies
  \begin{enumerate}[i.]
  \item Probability constraint (P):

    For any $\phi\subset\mathcal N$ and any $\mathbf t\in T$,
    \begin{equation}
      \label{p}
      \sum_{\psi\subset\mathcal N}p_\psi(\mathbf t) = 1\quad\textmd{and}\quad
      p_\phi(\mathbf t) \ge 0\ .
    \end{equation}
  \item Capacity constraint (CC):

    For any $i\in\mathcal N$, any $\mathbf t\in T$, and any
    $\phi\subset\mathcal N$ with $p_\phi(\mathbf t) > 0$,
    \begin{equation}
      \label{cc}
      \sum_{j\in\phi}x_{j, \phi}(\mathbf t) \le
      1\quad\textmd{and}\quad x_{i, \phi}(\mathbf t) \ge 0\ .
    \end{equation}
  \item Individual rationality (IR):

    For any $\phi\subset\mathcal N$ with $p_\phi(\mathbf t) > 0$, any
    $\mathbf t\in T$, and any $i\in\phi$,
    \begin{equation}
      \label{ir}
      x_{i, \phi}(\mathbf t) \ge t_i\ .
    \end{equation}
  \item Incentive compatibility (IC) in (\ref{ic}).
  \end{enumerate}
\end{mydef}

\begin{mydef}[Weak-determinism - weak-D]
  \label{weakdet}
  A mechanism $(p, x)$ is \emph{weakly-deterministic} if for any
  $\mathbf t\in T$ and any $i\in\mathcal N$,
  \begin{equation*}
    (\forall j\in\mathcal N, j\neq i\Rightarrow t_j\neq
    t_i)\Rightarrow\sum_{\phi:i\in\phi}p_\phi(\mathbf t)\in\{0, 1\}\ .
  \end{equation*}
  That is, for a given bid vector $\mathbf t$, if user $i$ is the only
  one who bids $t_i$, then either the user always gets admitted, or
  the user never gets admitted.
\end{mydef}
\begin{mydef}[Determinism - D]
  \label{det}
  A mechanism $(p, x)$ is \emph{deterministic} if there exists a
  function $\psi: T\rightarrow\mathcal P(\mathcal N)$ such that for
  any $\mathbf t\in T$,
  \begin{equation*}
    p_\phi(\mathbf t) = \begin{cases}
      1 & \textmd{if } \phi = \psi(\mathbf t)\\
      0 & \textmd{otherwise}
    \end{cases}
  \end{equation*}
  where $\mathcal P(A)$ is the power set of $A$. Determinism implies
  that the winning set is always unique. We call $\psi(\cdot)$ the
  \emph{winning set function}.
\end{mydef}

Note that for a deterministic mechanism $(p, x)$ we can have the
assigned service rate denoted as
\begin{equation}
  \label{singlesub}
  x_i(\mathbf t) = \begin{cases}
    x_{i, \psi(t)}(\mathbf t) & \textmd{if }i\in\psi(\mathbf t)\\
    0 & \textmd{otherwise}
  \end{cases}\ .
\end{equation}
\begin{mydef}[Anonymity]
  A mechanism $(p, x)$ is \emph{anonymous} if for any $\mathbf t\in
  T$, any $\pi\in\Gamma_n$, and any $\phi\subset\mathcal N$,
  \begin{align*}
    p(\pi(\mathbf t)) & = \pi(p(\mathbf t))\\
    x_\phi(\pi(\mathbf t)) & = \pi(x_\phi(\mathbf t)),
  \end{align*}
  where $\Gamma_n$ is the set of all permutations of $n$ indices. That
  is, the outcome of the auction does not depend on the identity of
  the bidders.
\end{mydef}
\begin{mydef}[Monotonicity]
  A mechanism $(p, x)$ is \emph{monotonic} if for any $i\in\mathcal
  N$, any $\mathbf t_{-i}\in T_{-i}$, and any $s_i, s_i'\in T_i$ with
  $s_i > s_i'$,
  \begin{equation*}
    \sum_{\phi: i\in\phi}p_\phi(\mathbf t_{-i}, s_i) \le \sum_{\phi:
      i\in\phi}p_\phi(\mathbf t_{-i}, s_i')\ .
  \end{equation*}
  That is, given that others' bids are fixed, a user's chance of
  getting admitted should not decrease when the user bids lower.
\end{mydef}

We have the following lemma for monotonicity.
\begin{mylemma}
  \label{icmon}
  Any mechanism that satisfies P, IR and IC is monotonic.
\end{mylemma}

\begin{JSAC}
	Due to page limitations, the proof of Lemma~\ref{icmon} is omitted. The interested reader can find it in \cite{kjy}.
\end{JSAC}
\begin{ARXIV}
	The proof of Lemma~\ref{icmon} is deferred to Appendix~\ref{pficmon}.
\end{ARXIV}
The idea of the proof is that if the mechanism is not monotonic, then there
exists a utility function such that users have incentive to lie.

We are interested in feasible auction mechanisms that are
weakly-deterministic and anonymous. Feasibility implies that the
decision is in the capacity region, no user is forced to participate,
and no one has incentive to lie about his type. Weak-determinism,
anonymity and monotonicity are properties we consider desirable for a
fair mechanism.
\begin{mydef}[Single-price]
  A mechanism $\mathcal A = (p, x)$ is \emph{single-priced} if for any
  $\mathbf t\in T$ and any $\phi\subset\mathcal N$ with
  $p_\phi(\mathbf t) > 0$,
  \[x_{i, \phi}(\mathbf t) = x_{j, \phi}(\mathbf t)\quad\forall i,
  j\in\phi\ .\] That is, all the users in a winning set get the same
  assigned service rate.
\end{mydef}

\section{Analysis of the Problem}
\label{analysis}
In this section we analyze feasible strategy-proof mechanisms in
non-monetary scenarios. We start with a mechanism $\bar{\mathcal A} =
(\bar p, \bar x)$, which is inspired by the uniform-price auction
\cite{kri}.
\subsection{The Simple Single-Priced Mechanism $\bar{\mathcal A}$}
We now propose the mechanism $\bar{\mathcal A}$, and prove that this
mechanism satisfies the capacity constraint, is weak-IC, and admits at least half of the maximum possible
number of users. After that we point out the flaws of the weak-IC
concept.

Given bid vector $\mathbf t\in T$, the mechanism 
$\bar{\mathcal A}$ is described as follows.
\begin{enumerate}[Step i.]
\item Let $\alpha$ be a rearrangement of the indices such that
  $t_{\alpha(1)} \le t_{\alpha(2)} \le \cdots \le t_{\alpha(n)}$. If
  several users bid the same, just arrange them randomly. Introduce a
  pseudo-bidder with $t_{n+1} = 1$ and $\alpha(n+1) = n+1$.
\item Find the largest index $\bar m$ with $1 \le \bar m \le n$ such
  that $\bar m\cdot t_{\alpha(\bar m+1)} \le 1$. The winning set is
  $\bar\psi(t) = \{\alpha(1), \alpha(2), \cdots, \alpha(\bar m)\}$.
\item Set the assigned service rate $\bar x_i(\mathbf t) = \bar x(\mathbf t) =
  t_{\alpha(\bar m+1)}$ if $i\in\bar\psi(\mathbf t)$.
\end{enumerate}

The idea of the mechanism is basically that we start from the lower
bidders and try to admit as many users as possible, with assigned service rate equal
to the lowest losing bid.

Notice that the rearrangement $\alpha$ and the winning set $\bar\psi$
here might not be unique if several users bid the same value, in which
case both of them will be probabilistic functions instead of
deterministic functions. However, when there are no two users bidding
the same value, $\alpha$ and $\bar\psi$ are deterministic functions.

Given the bids $\mathbf t\in T$ and a corresponding rearrangement
$\alpha$, we let
\begin{equation*}
  \sigma_i(\mathbf t) = t_{\alpha(i)},\quad 1 \le i \le n+1
\end{equation*}
and
\begin{equation*}
  \sigma(\mathbf t) = (\sigma_1(\mathbf t), \sigma_2(\mathbf t), \cdots,
  \sigma_{n+1}(\mathbf t))
\end{equation*}
where $\sigma(\mathbf t)$ is the unique sorted vector of $\mathbf t$
with $\sigma_1(\mathbf t) \le \sigma_2(\mathbf t) \le \cdots \le
\sigma_{n+1}(\mathbf t)$, regardless of the possible different choices
of $\alpha$. Then the index chosen in step ii is a function of the
bids given by
\begin{equation*}
  \bar m(\mathbf t) = \max\{m\in\mathcal
  N|m\cdot\sigma_{m+1}(\mathbf t) \le 1\}
\end{equation*}
where $\bar m(\mathbf t)$ is always well-defined because
\begin{equation*}
  1\cdot t_{\alpha(2)} \le 1\ .
\end{equation*}
Similarly, the assigned service rate for those admitted users is also determined by
\begin{equation*}
  \bar x(\mathbf t) = t_{\alpha(\bar m(\mathbf t)+1)}\ .
\end{equation*}

\begin{myexample}
  \label{eg*}
  Let the bid vector be $\mathbf t = (t_1, t_2, t_3, t_4) = \{0.5,
  0.4, 0.3, 0.4\}$.
  \begin{enumerate}[Step i.]
  \item The rearrangement could be $\alpha = (\alpha(1), \alpha(2),
    \alpha(3), \alpha(4), \alpha(5)) = (3, 2, 4, 1, 5)$. ($\alpha$
    could also be $(3, 4, 2, 1, 5)$) So $t_{\alpha(1)} \le
    t_{\alpha(2)} \le t_{\alpha(3)} \le t_{\alpha(4)} \le
    t_{\alpha(5)} = 1$ and the sorted vector of $\mathbf t$ is
    $\sigma(\mathbf t) = (0.3, 0.4, 0.4, 0.5, 1)$.
  \item $2\times 0.4 = 0.8 \le 1$ and $3\times 0.5 = 1.5 > 1$, so
    $\bar m(\mathbf t) = 2$ and the winning set is $\bar\psi(\mathbf
    t) = \{2, 3\}$ since $\alpha = (3, 2, 4, 1,
    5)$. ($\bar\psi(\mathbf t)$ would be $\{3, 4\}$ if $\alpha = (3,
    4, 2, 1, 5)$.)
  \item The assigned service rate for either of the two winners is $\bar x(\mathbf
    t) = t_{\alpha(3)} = t_4 = 0.4$. (Note that if $\alpha = (3, 4, 2,
    1, 5)$, then $\bar x(\mathbf t)$ would still be $0.4$.)
  \end{enumerate}
\end{myexample}
\begin{mylemma}
  \label{feasibility}
  The mechanism $\bar{\mathcal A}$ satisfies the P, CC, IR and weak-IC
  constraints.
\end{mylemma}

\begin{JSAC}
	Due to page limitations, the proof of Lemma~\ref{feasibility} is omitted. The interested reader can find it in \cite{kjy}.
\end{JSAC}
\begin{ARXIV}
	The proof of Lemma~\ref{feasibility} is deferred to Appendix~\ref{pffea}.
\end{ARXIV}
It follows from directly checking the P, CC,
and IR constraints. To verify weak-IC, we focus on the set of bid vectors
with no equal bids from any two users.

Note that $\bar{\mathcal A}$ does not satisfy
feasibility because it is not IC. To see this, just consider two users
bidding the same service rate. The chance of getting admitted is half
for either user. The chance of either user getting admitted increases
to 1 when he lower his bid by a small amount and the other user keeps
the original bid. $\bar{\mathcal A}$, however, is weakly-IC. More
specifically, for any $\mathbf t$ such that all bids are distinct,
$\bar{\mathcal A}$ guarantees truth-telling.

We now show that $\bar{\mathcal A}$ admits at least half of the
maximal possible number of users.
\begin{mythm}[Scalability of $\bar{\mathcal A}$]
  \label{scalability}
  For any true value of drop rate $\mathbf t\in T$, if there exists some
  mechanism with P, CC and IR that admits $m$ users, then
  $\bar{\mathcal A}$ can admit at least $\lfloor\frac{m}{2}\rfloor$
  users.
\end{mythm}

\begin{JSAC}
	Due to page limitations, the proof of Theorem~\ref{scalability} is omitted. The interested reader can find it in \cite{kjy}.
\end{JSAC}
\begin{ARXIV}
	The proof of Theorem~\ref{scalability} is deferred to Appendix~\ref{pfsca}.
\end{ARXIV}
The intuition behind it is that any algorithm that
tries to admit more than $m$ users will violate CC.

The problem about $\bar{\mathcal A}$ is that it is only weakly-IC but
not IC. This means that if equal bids exist, users might have
incentive to lie. For example, in a two-user case, if both users bid
the same value, each of them would have half chance of getting
admitted. But if one of them lower his bid by a small amount, he would
win with the same assigned service rate and probability 1. Thus indistinguishable
bids make $\bar{\mathcal A}$ fail for IC.
\subsection{Impossibility for Probabilistic Decisions of Equal
  Bids}
We now show that to fulfill strict IC under some assumptions mentioned
below, a weak-deterministic mechanism has to be deterministic. That
is, if several users bid exactly the same value, then the only choice
for guaranteeing truth-telling is to either admit all or none of them.
\begin{mythm}[Impossibility]
  \label{impossibility}
  For a mechanism $\mathcal A$ that satisfies P, IR, IC and anonymity,
  $\mathcal A$ is weakly-deterministic if and only if $\mathcal A$ is
  deterministic. That is, $\mathcal A$ admits either all or none of
  the equal bids.
\end{mythm}

\begin{JSAC}
	Due to page limitations, the proof of Theorem~\ref{impossibility} is omitted. The interested reader can find it in \cite{kjy}.
\end{JSAC}
\begin{ARXIV}
	The proof of Theorem~\ref{impossibility} is deferred to Appendix~\ref{pfimp}.
\end{ARXIV}
By definition we know that determinism implies
weak-determinism, so we only need to prove that weak-determinism
implies determinism. To do that, we prove by contradiction that if an auction
is weakly-deterministic and not deterministic, then there exists utility functions
such that users have an incentive to lie.

This theorem indicates that for any feasible and
weakly-deterministic scheme, the ratio between the maximal possible
number of users and the number of users admitted under the scheme is
unbounded.

We should note that $\bar{\mathcal A}$ is weakly-deterministic because
users might be randomly admitted when bidding the same. Then by
Theorem~\ref{impossibility}, to achieve IC we need to design
deterministic mechanisms, that is, mechanisms with only deterministic
outcomes.

In the next section, we will present a feasible scheme based on
$\bar{\mathcal A}$. Before that, we first present some properties any
feasible scheme should have.
\subsection{Highest Winning Bid Theorem}
We further show that any deterministic mechanism with IC must be
illustrated by a \emph{highest winning bid function}.
\begin{mydef}[Highest winning bid mechanism]
  \label{lwbdef}
  A deterministic mechanism $\mathcal A$ is a \emph{highest
    winning bid mechanism} if there exists some function $z\colon
  T_{-i}\to T_i$ such that for any $\mathbf t_{-i}\in T_{-i}$,
  \begin{equation*}
    \label{lwb}
    \begin{cases}
      \textmd{if }s_i \le z(\mathbf t_{-i}) & \textmd{then
      }i\in\psi(\mathbf t_{-i}, s_i)\textmd{ and }x_i(\mathbf t_{-i},
      s_i) = z(\mathbf t_{-i})\\
      \textmd{if }s_i > z(\mathbf t_{-i}) & \textmd{then }
      i\notin\psi(\mathbf t_{-i}, s_i)\textmd{ and }x_i(\mathbf
      t_{-i}, s_i) = 0
    \end{cases}
  \end{equation*}
  where $\psi(\cdot)$ is the winning set function of $\mathcal A$ defined
  in the definition of determinism (Definition~\ref{det}).

  The function $z(\cdot)$ is called the \emph{highest winning bid
    function} of $\mathcal A$.
\end{mydef}
\begin{mythm}[Highest winning bid]
  \label{lwbthm}
  A deterministic mechanism satisfies IC if and only if it is a
  highest winning bid mechanism.
\end{mythm}

\begin{JSAC}
	Due to page limitations, the proof of Theorem~\ref{lwbthm} is omitted. The interested reader can find it in \cite{kjy}.
\end{JSAC}
\begin{ARXIV}
	The proof of Theorem~\ref{lwbthm} is deferred to Appendix~\ref{pflwbthm}.
\end{ARXIV}
It can be checked that a highest
winning mechanism satisfies IC, so we only need to prove the converse.
To do that, we show that if a deterministic mechanism satisfies IC, then
we can always construct a highest winning bid function.

The highest winning bid theorem shows us what a deterministic
truth-telling mechanism should look like. Notice that no other
assumptions are needed for this theorem, so it remains valid in a
general setting. More importantly, this theorem gives us an efficient
approach to design deterministic truth-telling mechanisms.

Although Theorem~\ref{lwbthm} does not work for $\bar{\mathcal A}$ due
to weak-determinism, we do have the following similar result.
\begin{mylemma}[Supremum winning bid function for $\bar{\mathcal A}$]
  \label{iwbprop}
  Under mechanism $\bar{\mathcal A}$, knowing others' bid $\mathbf
  t_{-i}$, the supremum of user $i$'s winning bids is given by
  \begin{equation}
    \label{iwbfunc}
    \bar z(\mathbf t_{-i}) = \max\{\sigma_j(\mathbf
    t_{-i})|j\sigma_j(\mathbf t_{-i}) \le 1\}\ .
  \end{equation}
\end{mylemma}

\begin{JSAC}
	Due to page limitations, the proof of Lemma~\ref{iwbprop} is omitted. The interested reader can find it in \cite{kjy}.
\end{JSAC}
\begin{ARXIV}
	The proof of Lemma~\ref{iwbprop} is deferred to Appendix~\ref{pfiwbprop}.
\end{ARXIV}
It follows from checking that under $\bar{\mathcal A}$,
and for all $i$, no bid larger than $\bar z(\mathbf t_{-i})$ is admitted.

Note that $\bar z(\mathbf t_{-i})$ is not the highest winning bid for
user $i$ because bidding this value does not guarantee winning.
\section{Our Proposed Mechanism}
\label{mechanism}
We now introduce the mechanism $\bar{\mathcal A}^*$, which is a
truth-telling mechanism based on the previous $\bar{\mathcal A}$. We
first construct $\bar{\mathcal A}^*$ by the so-called \emph{dropping
  trick}. After that we show that $\bar{\mathcal A}^*$ is
single-priced, feasible, and has very similar behavior to
$\bar{\mathcal A}$.
\subsection{Dropping Trick}
The basic idea of the dropping trick is that since
bidding exactly the supremum winning bid does not guarantee winning
due to the capacity constraint, we drop the function by a small amount
whenever necessary such that the capacity constraint is satisfied.

We would like to find a highest winning bid function $\bar z^*\colon
T_{-i}\to T_i$ based on the supremum winning bid function $\bar
z(\cdot)$ of $\bar{\mathcal A}$. For any $\mathbf t_{-i}\in T_{-i}$,
let $\sigma_0(\mathbf t_{-i}) = 0$, $\sigma_n(\mathbf t_{-i}) = 1$,
and let
\[m_1(\mathbf t_{-i}) = \max_j\{j|j\sigma_j(\mathbf t_{-i}) \le 1\}\]
and
\[m_2(\mathbf t_{-i}) = \max_j\{j|(j+1)\sigma_j(\mathbf t_{-i}) \le
1\}\ .\] Then, we can also write the supremum winning bid function for
$\bar{\mathcal A}$ as follows
\[\bar z(\mathbf t_{-i}) = \sigma_{m_1(\mathbf t_{-i})}(\mathbf
t_{-i})\ .\]

It must be noted that from the definition of $m_1(\mathbf t_{-i})$
and $m_2(\mathbf t_{-i})$, whenever $m_1(\mathbf t_{-i}) \neq m_2(\mathbf t_{-i})$,
it must be the case that $m_2(\mathbf t_{-i}) < m_1(\mathbf t_{-i})$. Thus,
if $m_1(\mathbf t_{-i}) \neq m_2(\mathbf t_{-i})$ we have
from the definition of $m_2(\mathbf t_{-i})$ that
\[(m_1(\mathbf t_{-i})+1) \sigma_{m_1(\mathbf t_{-i})}(\mathbf
t_{-i}) >1,\]
or equivalently,

\[  \bar z(\mathbf t_{-i}) = \sigma_{m_1(\mathbf t_{-i})}(\mathbf t_{-i}) > \frac{1}{ m_1(\mathbf t_{-i})+1 }. \]

For fixed parameters $(d_j, 1\le j\le n-1)$ with $0 < d_j < 1$, the
highest winning bid function of $\bar{\mathcal A}^*$ is defined as
follows:
\begin{equation}
  \label{lwbfunc}
  \bar z^*(\mathbf t_{-i}) =
  \begin{cases}
    \bar z(\mathbf t_{-i}) & \textmd{if }m_1(\mathbf t_{-i}) =
    m_2(\mathbf t_{-i})\\
    \bar z(\mathbf t_{-i}) \left( 1-d_{m_1(\mathbf t_{-i})} \right) + \frac 1{m_1(\mathbf t_{-i})+1} d_{m_1(\mathbf t_{-i})} & \textmd{if
    }m_1(\mathbf t_{-i}) \neq m_2(\mathbf t_{-i})
  \end{cases}
\end{equation}
or equivalently
\begin{equation*}
  \label{lwbfunc2}
  \bar z^*(\mathbf t_{-i}) =
  \begin{cases}
    \sigma_{m_1(\mathbf t_{-i})}(\mathbf t_{-i}) & \textmd{if
    }\sigma_{m_1(\mathbf t_{-i})}(\mathbf t_{-i}) \le
    \frac{1}{m_1(\mathbf t_{-i})+1}\\
    \sigma_{m_1(\mathbf t_{-i})}(\mathbf t_{-i}) \left( 1-d_{m_1(\mathbf t_{-i})} \right) +\frac
      1{m_1(\mathbf t_{-i})+1}d_{m_1(\mathbf
      t_{-i})} & \textmd{if
    }\sigma_{m_1(\mathbf t_{-i})}(\mathbf t_{-i}) >
    \frac{1}{m_1(\mathbf t_{-i})+1}\ .
  \end{cases}
\end{equation*}
Thus, the parameter $d_{m_1(\mathbf t_{-i})}$ \emph{drops} the value of $\bar z^*(\mathbf t_{-i})$
to lay on the interval $\left( \frac 1{m_1(\mathbf t_{-i})+1} , \bar z(\mathbf t_{-i}) \right)$
whenever $m_1(\mathbf t_{-i}) \neq m_2(\mathbf t_{-i})$.

Note that, compared to the supremum winning bid function of
$\bar{\mathcal A}$, we only do dropping when $m_1(\mathbf t_{-i}) \neq
m_2(\mathbf t_{-i})$.  We note that $\bar{\mathcal A}^*$ is a
deterministic mechanism based on the highest winning bid function
$\bar z^*(\cdot)$ in (\ref{lwbfunc}) and by Theorem~\ref{lwbthm} we
know that $\bar{\mathcal A}^*$ satisfies IC.

We would like to highlight the fact that neither $\bar{\mathcal A}$ nor $\bar{\mathcal
  A}^*$ dominate the other in terms of maximizing the number of
admitted users. That is, for some bid vectors $\bar{\mathcal A}$
admits more users than $\bar{\mathcal A}^*$ does and for some others
$\bar{\mathcal A}^*$ admits more than $\bar{\mathcal A}$ does. For
example, consider the 3-user case and let the parameters for
$\bar{\mathcal A}^*$ be $d_1 = d_2 = 0.1$. For bid vector $(0.79, 0.8,
0.9)$, the first user would be admitted by $\bar{\mathcal A}$ with service rate 0.8, but it
would not be admitted by $\bar{\mathcal A}^*$ since $\bar z^*(0.8,
0.9) = 0.8-0.1\times(0.8-0.5) = 0.77 < 0.79$. Following a similar analysis for bid vector $(0.1,
0.1, 0.9)$, we note that $\bar{\mathcal A}$ only admits one user
since it cannot admit two users with service rate 0.9, 
while $\bar{\mathcal A}^*$ can admit two users since $\bar z^*(0.1, 0.9) =
0.1$.
\begin{myexample}
  Take $d_j = 0.1$ for any $1 \le j \le n-1$ in
  (\ref{lwbfunc}). Again, let the bid vector be $\mathbf t = (t_1,
  t_2, t_3, t_4) = \{0.5, 0.4, 0.3, 0.4\}$. By (\ref{iwbfunc}) we can
  calculate the supremum winning bid under $\bar{\mathcal A}$ for each
  user:
  \begin{align*}
    \bar z(\mathbf t_{-1}) & = \bar z(0.4, 0.3, 0.4) = 0.4\ ,\\
    \bar z(\mathbf t_{-2}) & = \bar z(0.5, 0.3, 0.4) = 0.4\ ,\\
    \bar z(\mathbf t_{-3}) & = \bar z(0.5, 0.4, 0.4) = 0.4\ ,\\
    \bar z(\mathbf t_{-4}) & = \bar z(0.5, 0.4, 0.3) = 0.4\ .
  \end{align*}
  Then the highest winning bid under $\bar{\mathcal A}^*$ for each
  user is
  \begin{align*}
    \bar z^*(\mathbf t_{-1}) & = \bar z(\mathbf t_{-1})-d_2\left(\bar
      z(\mathbf t_{-1})-\frac 13\right) \doteq 0.3933\ ,\\
    \bar z^*(\mathbf t_{-2}) & = \bar z(\mathbf t_{-2})-d_2\left(\bar
      z(\mathbf t_{-2})-\frac 13\right) \doteq 0.3933\ ,\\
    \bar z^*(\mathbf t_{-3}) & = \bar z(\mathbf t_{-3})-d_2\left(\bar
      z(\mathbf t_{-3})-\frac 13\right) \doteq 0.3933\ ,\\
    \bar z^*(\mathbf t_{-4}) & = \bar z(\mathbf t_{-4})-d_2\left(\bar
      z(\mathbf t_{-4})-\frac 13\right) \doteq 0.3933\ .
  \end{align*}
  Since only user 3's bid is lower than or equal to his highest
  winning bid, we have the winning set $\bar\psi^*(\mathbf t) = \{3\}$
  and $\bar x_3^*(\mathbf t) = 0.3933$.
\end{myexample}
\subsection{Properties of $\bar{\mathcal A}^*$}
We first notice that the dropping method above is chosen such that the
mechanism remains single-priced.
\begin{mylemma}
  \label{sp}
  $\bar{\mathcal A}^*$ is single-priced.
\end{mylemma}

\begin{JSAC}
	Due to page limitations, the proof of Lemma~\ref{sp} is omitted. The interested reader can find it in \cite{kjy}.
\end{JSAC}
\begin{ARXIV}
	The proof of Lemma~\ref{sp} is deferred to Appendix~\ref{pfsp}.
\end{ARXIV}

We then notice that the mechanism $\bar{\mathcal A}^*$ is indeed
feasible.
\begin{mylemma}
  \label{feasibility2}
  $\bar{\mathcal A}^*$ is feasible, and $0 \le \bar z^*(\mathbf
  t_{-i}) \le 1$ for any $\mathbf t_{-i}\in T_{-i}$.
\end{mylemma}

\begin{JSAC}
	Due to page limitations, the proof of Lemma~\ref{feasibility2} is omitted. The interested reader can find it in \cite{kjy}.
\end{JSAC}
\begin{ARXIV}
	The proof of Lemma~\ref{feasibility2} is deferred to Appendix~\ref{pffea2}.
\end{ARXIV}
The key aspect of the proof is to show that the
CC constraint is fulfilled. To do that, the proof considers the case when
the dropping trick is used and when it is not. From that, and from the analysis
of $\bar{\mathcal A}$, the capacity constraint can be verified.
\subsection{Performance Analysis}
We now have two single-priced mechanisms and would like to compare
these to some optimal, non-truthful mechanisms that
maximize the number of admitted users. We first introduce two
optimal omniscient auctions \cite{ghkws}.
\begin{mydef}
  Given bid vector $\mathbf t$, the \emph{optimal single price
    omniscient auction} $\mathcal F$ admits the lowest $m_{\mathcal
    F}^*$ users with
  \[m_{\mathcal F}^* = \max\{m|m\sigma_m(\mathbf t) \le 1\}\ .\]
  Compared to $\bar{\mathcal A}$, $\mathcal F$ use the highest winning
  bid as the universal price instead of the lowest losing
  bid. It must be noted though that $\mathcal F$ is non-deterministic and non-truthful.
\end{mydef}
\begin{mydef}
  Given bid vector $\mathbf t$, the \emph{optimal multiple price
    omniscient auction} $\mathcal T$ admits the lowest $m_{\mathcal
    T}^*$ users with each winner's price equal to his own bid.
\end{mydef}
It is easy to see that $\mathcal T$ is not single-priced, non-deterministic
and non-truthful, and admits the maximum number of users. To sum up,
we list all the mechanisms we want to compare and their corresponding properties
in Table~\ref{tb:1}.
\begin{table}[h!]
  \centering
  \begin{tabular}{cccccc}
    & Optimal & Deterministic & Single Price & Weak-IC & IC\\
    $\mathcal F$ & \tick & \cross & \tick & \cross &
    \cross\\
    $\mathcal T$ & \tick & \cross & \cross & \cross &
    \cross\\
    $\bar{\mathcal A}$ & \cross & \cross & \tick & \tick &
    \cross\\
    $\bar{\mathcal A}^*$ & \cross & \tick & \tick & \tick &
    \tick
  \end{tabular}
  \caption{Comparison of the Mechanisms}
  \label{tb:1}
\end{table}
\begin{mydef}
  Given bid vector $\mathbf t$, the \emph{admittance} of a mechanism
  $\mathcal A$ is the expected number of users admitted by $\mathcal
  A$, and we denote it by $|\mathcal A(\mathbf t)|$.
\end{mydef}
We consider
both Bayesian analysis and worst case analysis as follows.
\subsubsection{Worst Case Analysis}
First, by the scalability of $\bar{\mathcal A}$ we have
\[|\bar{\mathcal A}(\mathbf t)| \ge \left\lfloor\frac{|\mathcal
    T(\mathbf t)|}{2}\right\rfloor\] for any $\mathbf t\in T$. This
bound is tight since we can consider the bid vector with $m+1$ users
bidding $0$ and $m$ users bidding $\frac 1{m}$, in which case
$|\bar{\mathcal A}(\mathbf t)| = m$ and $|\mathcal T(\mathbf t)| =
2m+1$.

For single price mechanisms, we have the following worst case results.
\begin{mylemma}
  \label{wc1}
  For any $\mathbf t\in T$,
  \[|\mathcal F(\mathbf t)|-1 \le |\bar{\mathcal A}(\mathbf t)| \le
  |\mathcal F(\mathbf t)|\ .\]
\end{mylemma}
\begin{JSAC}
	Due to page limitations, the proof of Lemma~\ref{wc1} is omitted. The interested reader can find it in \cite{kjy}.
\end{JSAC}
\begin{ARXIV}
	The proof of Lemma~\ref{wc1} is deferred to Appendix~\ref{pfwc1}.
\end{ARXIV}
It follows
from the definition of $\mathcal F(\mathbf t)$ and $\bar{\mathcal A}(\mathbf t)$. It is
interesting to highlight that the bounds are tight since they are achievable.
\begin{mylemma}
  \label{wc2}
  For any $n\in\mathbb N$ and $m < n$, there exists some bid vector
  $\mathbf t$ such that
  \[|\bar{\mathcal A}^*(\mathbf t)| = 0\ ,\quad|\mathcal F(\mathbf t)|
  \ge m\ .\] That is to say, the worst case performance of
  $\bar{\mathcal A}^*$ could be arbitrarily bad.
\end{mylemma}
\begin{JSAC}
	Due to page limitations, the proof of Lemma~\ref{wc2} is omitted. The interested reader can find it in \cite{kjy}.
\end{JSAC}
\begin{ARXIV}
	The proof of Lemma~\ref{wc2} is deferred to Appendix~\ref{pfwc2}.
\end{ARXIV}
It uses the fact that
in $\bar{\mathcal A}^*(\mathbf t)$ all
identical bids must be either accepted or rejected, while this is not
the case for $\mathcal F(\mathbf t)$. Note
that while the worst case performance of $\bar{\mathcal A}^*$ can be
bad, we will next show that the probability of getting a worst case
can be made very small by setting small parameters for $\bar{\mathcal
  A}^*$.

\subsubsection{Bayesian Analysis}
We have shown that $\bar{\mathcal A}$ has roughly the same performance
than $\mathcal F$ in terms of maximizing the number of admitted
users.

Now we show that $\bar{\mathcal A}^*$ has very close performance to
$\bar{\mathcal A}$. Assume that the drop rate vector $t$ is drawn from
a distribution with joint probability density function $f\colon
T\to\mathbb R^+$. Then, we have the following theorem.
\begin{mythm}
  \label{bp}
  If the density function $f$ is upper-bounded by $K$, the probability that
  $\bar{\mathcal A}^*$ behaves differently from $\bar{\mathcal A}$ is
  at most $dnK$, where $d = \max_{1\le j\le n-1}d_j$.
\end{mythm}

\begin{JSAC}
	Due to page limitations, the proof of Theorem~\ref{bp} is omitted. The interested reader can find it in \cite{kjy}.
\end{JSAC}
\begin{ARXIV}
	The proof of Theorem~\ref{bp} is deferred to Appendix~\ref{pfbp}.
\end{ARXIV}
It is
based on the fact that if the density function is bounded, then the problem
simplifies to bounding the Lebesgue measure of the set of bids where
$\bar{\mathcal A}^*$ behaves differently from $\bar{\mathcal A}$.

We notice from Theorem~\ref{bp} that the probability of different
behaviors between the two mechanisms can be made arbitrarily small by choosing
small parameters $(d_j, 1\le j\le n-1)$.

\section{Conclusion}
\label{conclusion}
In this paper, we studied the problem of designing a strategy-proof
non-monetary auction mechanism for wireless networks. The motivation
is to let the users tell the truth when bidding their resource
requirements, and to admit as many users as possible. We gave a
general model for the problem, analyzed the problem and
found some properties that any strategy-proof auction mechanism should
satisfy. Finally we proposed a feasible mechanism which is truthful
even with equal bids, and showed that it could admit at least half of
the maximum number of users with high probability in an asymptotic
sense.

As possible topic for future work, discrete pricing models might be
considered rather than continuous pricing models. Also, the assumption
of weak-determinism could be weakened, and more specific utility
functions could be considered for better performance. Furthermore, the
lower bound of the number of admitted users might be improved.

\bibliographystyle{IEEEtran}
\bibliography{IEEEabrv,auction}
\begin{ARXIV}
\section{Appendix}
\subsection{Proof of Lemma~\ref{icmon}}
\label{pficmon}
\begin{IEEEproof}
  Let mechanism $(p, x)$ satisfy P, IR, IC. Suppose $(p, x)$ is not
  monotonic, then there exist $i\in\mathcal N$, $\mathbf t_{-i}\in
  T_{-i}$, and $s_i, s_i'\in T_i$ such that $s_i > s_i'$ and
  \begin{equation}
    \label{pficmon:eq1}
    \sum_{\phi: i\in\phi}p_\phi(\mathbf t_{-i}, s_i) >
    \sum_{\phi: i\in\phi}p_\phi(\mathbf t_{-i}, s_i')\ .
  \end{equation}
  We consider the following utility function:
  \begin{equation*}
    u_i(t_i, x_i) =
    \begin{cases}
      d & \textmd{if } x_i \ge t_i\\
      0 & \textmd{if } x_i < t_i
    \end{cases},
  \end{equation*}
  where $d$ is a positive constant. Let the true service rate value of
  user $i$ be $q_i = s_i'$, then
  \begin{align}
    U(p, x, q_i, s_i, \mathbf t_{-i}) & = U(p, x, s_i', s_i,
    \mathbf t_{-i})\nonumber\\
    & = \sum_{\phi:i\in\phi}u_i(s_i', x_{i, \phi}(\mathbf t_{-i},
    s_i))p_\phi(\mathbf t_{-i}, s_i)\nonumber\\
    \label{pficmon:eq2}
    & = d\sum_{\phi:i\in\phi}p_\phi(\mathbf t_{-i}, s_i)\\
    \label{pficmon:eq3}
    & > d\sum_{\phi:i\in\phi}p_\phi(\mathbf t_{-i}, s_i')\\
    \label{pficmon:eq4}
    & = \sum_{\phi:i\in\phi}u_i(s_i', x_{i, \phi}(\mathbf t_{-i},
    s_i'))p_\phi(\mathbf t_{-i}, s_i')\\
    & = U(p, x, s_i', s_i', \mathbf t_{-i})\nonumber\\
    & = U(p, x, t_i, t_i, \mathbf t_{-i})\nonumber
  \end{align}
  where (\ref{pficmon:eq2}) comes from $x_{i, \phi}(\mathbf t_{-i},
  s_i) \ge s_i \ge s_i'$ when $p_\phi(\mathbf t_{-i}, s_i) > 0$,
  (\ref{pficmon:eq4}) comes from $x_{i, \phi}(\mathbf t_{-i}, s_i')
  \ge s_i'$ when $p_\phi(\mathbf t_{-i}, s_i') > 0$, and
  (\ref{pficmon:eq3}) comes from (\ref{pficmon:eq1}). This contradicts
  IC. Thus, $(p, x)$ must be monotonic.
\end{IEEEproof}
\subsection{Proof of Lemma~\ref{feasibility}}
\label{pffea}
\begin{IEEEproof}
  \begin{enumerate}[i.]
  \item Probability constraint (P):

    The probability constraint is obviously satisfied. For those
    $\mathbf t$ such that the winning set $\bar\psi(\mathbf t)$ is
    determined, $\sum_{\phi\subset\mathcal N}\bar p_\phi(\mathbf t) =
    \bar p_{\bar\psi(\mathbf t)}(\mathbf t) = 1$. For those $\mathbf
    t$ such that there are $M$ possible winning sets, the probability
    for each of them would be $\frac{1}{M}$ and
    $\sum_{\phi\subset\mathcal N}\bar p_\phi(\mathbf t) = M\cdot\frac{1}{M}
    = 1$.
  \item Capacity constraint (CC):

    For any $\mathbf t\in T$ and any $\phi\subset\mathcal N$ with
    $\bar p_\phi(\mathbf t) > 0$,
    \begin{align*}
      \sum_{i\in\phi}\bar x_{i, \phi}(\mathbf t) & =
      \bar m(\mathbf t)\cdot\bar x(\mathbf t)\\
      & = \bar m(\mathbf t)\cdot t_{\alpha(\bar m(\mathbf t)+1)}\\
      & \le \bar m(\mathbf t)\cdot\frac{1}{\bar m(\mathbf t)}\\
      & = 1\ .
    \end{align*}
    Also $\bar x_{i, \phi}(\mathbf t) \ge 0$. Thus, $\bar{\mathcal A}$
    satisfies CC.
  \item Individual rationality (IR):

    For any $\mathbf t\in T$, any $\phi\subset\mathcal N$ with
    $\bar p_\phi(\mathbf t) > 0$, and any $i\in\phi$,
    \begin{equation*}
      \bar x_{i, \phi}(\mathbf t) = \bar x(\mathbf t) =
      t_{\alpha(\bar m(\mathbf t)+1)} \ge t_i\ .
    \end{equation*}
  \item Weak-incentive compatibility (weak-IC):

    We only consider the set of \emph{distinguishable} bid vectors
    \begin{equation*}
      T_D = \{\mathbf t\in T|t_i\neq t_j \forall i\neq j\},
    \end{equation*}
    that is, the set of bid vectors with no equal bids from any two
    users. For $\mathbf t\in T_D$ and $i\in\mathcal N$, the result of
    user $i$ bidding $s_i$ would be
    \begin{equation*}
      \bar p_i(\mathbf t_{-i}, s_i) =
      \begin{cases}
        1 & \textmd{if }s_i < \bar x(\mathbf t)\\
        \frac{1}{2} & \textmd{if }s_i = \bar x(\mathbf t)\\
        0 & \textmd{if }s_i > \bar x(\mathbf t)
      \end{cases}
    \end{equation*}
    with assigned service rate $\bar x_i(\mathbf t_{-i}, s_i) = \bar x(\mathbf t)$
    if admitted. Note that $\bar p_i(\mathbf t_{-i}, s_i) =
    \frac{1}{2}$ when $s_i = \bar x(\mathbf t)$ since there is only
    one other user who bids $\bar x(\mathbf t)$.

    We first consider the case of $i\in\bar\psi(\mathbf t)$. We then
    have $t_i < \bar x(\mathbf t)$. If $s_i < \bar x(\mathbf t)$, then
    user $i$ still gets admitted with the same assigned service rate. If $s_i > \bar
    x(\mathbf t)$, then user $i$ gets rejected. If $s_i = \bar
    x(\mathbf t)$, then user $i$ either gets admitted with the same
    assigned service rate, or get rejected, both of which have probability 1/2. So
    user $i$ cannot get better utility in the first case.

    We then consider the case of $i\notin\bar\psi(\mathbf t)$. Now we
    have $t_i > \bar x(\mathbf t)$. If $s_i < \bar x(\mathbf t)$, then
    user $i$ gets admitted with assigned service rate $\bar x(\mathbf t)$ lower than
    true value $t_i$. If $s_i > \bar x(\mathbf t)$, the user $i$ still
    does not get admitted. If $s_i = \bar x(\mathbf t)$, then user $i$
    either gets admitted with assigned service rate too low to accept, or does not
    get admitted at all, both of which have probability 1/2. So user
    $i$ cannot get better utility in the second case.

    Thus, for any $\mathbf t\in T_D$, no user has incentive to lie. As
    $T\backslash T_D$ has measure zero, we have weak-IC.
  \end{enumerate}
\end{IEEEproof}
\subsection{Proof of Theorem~\ref{scalability}}
\label{pfsca}
\begin{IEEEproof}
  Suppose $\bar{\mathcal A}$ admits only $m$ users, that is,
  \begin{equation*}
    \bar m(\mathbf t) = m
  \end{equation*}
  and some other mechanism $\mathcal A = (p, x)$ has a chance of
  admitting at least $2m+2$ users given bid vector $\mathbf t$, that
  is,
  \begin{equation*}
    \exists\phi\subset\mathcal N, |\phi| \ge 2m+2, p_\phi(\mathbf
    t) > 0\ .
  \end{equation*} Then we have
  \begin{align}
    \label{pfsca:eq1}
    \sum_{i\in\phi}x_{i, \phi}(\mathbf t) & \ge \sum_{i\in\phi}t_i\\
    & \ge \sum_{i = 1}^{2m+2}\sigma_{i}(\mathbf t)\nonumber\\
    & \ge \sum_{i = m+2}^{2m+2}\sigma_{i}(\mathbf t)\nonumber\\
    & \ge (m+1)\sigma_{m+2}(\mathbf t)\nonumber\\
    \label{pfsca:eq2}
    & > 1
  \end{align}
  where the inequality (\ref{pfsca:eq1}) comes from IR and
  (\ref{pfsca:eq2}) comes from the definition of the mechanism.

  Thus $\mathcal A$ admits at most $2m+1$ users. This is equivalent to
  the statement that if some mechanism with P, CC and IR admits $m$
  users, $\bar{\mathcal A}$ can at least admit
  $\lfloor\frac{m}{2}\rfloor$ users.
\end{IEEEproof}
\subsection{Proof of Theorem~\ref{impossibility}}
\label{pfimp}
\begin{IEEEproof}
  We first prove the following lemma.
  \begin{mylemma}
    \label{uni}
    A mechanism $\mathcal A = (p, x)$ is deterministic if and only if
    for any $\mathbf t\in T$ and any $i\in\mathcal N$,
    \begin{equation}
      \label{lem1}
      \sum_{\phi:i\in\phi}p_\phi(\mathbf t)\in\{0, 1\}\ .
    \end{equation}
  \end{mylemma}
  \begin{IEEEproof}
    The determinism of $\mathcal A$ implies (\ref{lem1}) because we
    have $\sum_{\phi:i\in\phi}p_\phi(\mathbf t) = p_{\psi(t)}(\mathbf
    t) = 1$ for $i\in\psi(\mathbf t)$, and
    $\sum_{\phi:i\in\phi}p_\phi(\mathbf t) = 0$ for
    $i\notin\psi(\mathbf t)$.

    If we have (\ref{lem1}) for a mechanism $\mathcal A$, then we let
    \[\tilde\psi(\mathbf t) = \{i\in\mathcal
    N|\sum_{\phi:i\in\phi}p_\phi(\mathbf t) = 1\}\] and claim that
    $\mathcal A$ is deterministic with the winning set function
    $\tilde\psi(\mathbf t)$.

    Indeed, suppose there exists $\psi\subset\mathcal N$ such that
    $p_\psi(\mathbf t) > 0$ and $\tilde\psi(\mathbf t)\neq\psi$. Then
    there exists $i\in(\tilde\psi(\mathbf
    t)\backslash\psi)\cup(\psi\backslash\tilde\psi(\mathbf t))$. If
    $i\in\tilde\psi(\mathbf t)\backslash\psi$,
    \begin{align*}
      \sum_{\phi:i\in\phi}p_\phi(\mathbf t) & =
      \sum_{\substack{\phi:i\in\phi\\\phi\neq\psi}}p_\phi(\mathbf t)\\
      & \le 1-p_\psi(\mathbf t)\\
      & < 1
    \end{align*}
    which contradicts the fact that $i\in\tilde\psi(\mathbf t)$. If
    $i\in\psi\backslash\tilde\psi(\mathbf t)$,
    \begin{equation*}
      \sum_{\phi:i\in\phi}p_\phi(\mathbf t) \ge p_\psi(\mathbf t) > 0
    \end{equation*}
    which also contradicts the fact that $i\notin\tilde\psi(\mathbf
    t)$. So for any $\psi$, $p_\psi(\mathbf t) > 0$ implies
    $p_\psi(\mathbf t) = 1$ and $\psi = \tilde\psi(\mathbf t)$. Hence
    $\mathcal A$ is deterministic with winning set function
    $\tilde\psi(\mathbf t)$.
  \end{IEEEproof}

  By definition we readily have that determinism leads to
  weak-determinism.

  On the other hand, suppose $\mathcal A$ is weakly-deterministic but
  not deterministic. By Lemma~\ref{uni}, there exists $\mathbf r\in T$
  and $i\in\mathcal N$ such that
  \[0 < \sum_{\phi:i\in\phi}p_\phi(\mathbf r) < 1\ .\]

  We first show that for any $r_i' < r_i$, we have
  \[\sum_{\phi:i\in\phi}p_\phi(\mathbf r_{-i}, r_i') = 1\ .\]
  Indeed, let
  \[\tilde r_i =
  \begin{cases}
    \max_{j\in\mathcal N}\{r_j|r_j < r_i\} & \textmd{if } r_j <
    r_i\textmd{ for some }j\in\mathcal N\\
    0 & \textmd{otherwise}
  \end{cases}\ .
  \]
  Then for any $r_i'\in(\tilde r_i, r_i)$, $r_i'$ is a unique bid in
  $(\mathbf r_{-i}, r_i')$, so by monotonicity and weak-determinism we
  have $\sum_{\phi:i\in\phi}p_\phi(\mathbf r_{-i}, r_i') = 1$. Again
  by monotonicity we also have $\sum_{\phi:i\in\phi}p_\phi(\mathbf
  r_{-i}, r_i') = 1$ for $r_i' \le \tilde r_i$.

  Let
  \[\Psi_i(\mathbf r) = \{\phi\subset\mathcal N|i\in\phi,
  p_\phi(\mathbf r) > 0\}\ .\] Then $\Psi_i(\mathbf r)$ is the set of
  possible winning sets that includes user $i$, given bid vector
  $\mathbf r$.

  Now fix $r_i' < r_i$, there are two cases. One is that there exists
  $\psi\in\Psi_i(\mathbf r_{-i}, r_i')$ with $x_{i, \psi}(\mathbf
  r_{-i}, r_i') < r_i$, and the other one is that for any
  $\psi\in\Psi_i(\mathbf r_{-i}, r_i')$, $x_{i, \psi}(\mathbf r_{-i},
  r_i') \ge r_i$. We want to prove that both cases lead to
  contradiction.

  For the first case, there exists $\psi\in\Psi_i(\mathbf r_{-i}, r_i')$
  with $x_{i, \psi}(\mathbf r_{-i}, r_i') < r_i$. Fix $r_i''\in(x_{i,
    \psi}(\mathbf r_{-i}, r_i'), r_i)$. We consider the following
  utility function:
  \begin{equation}
    \label{uti1}
    u_i(t_i, x_i) = \begin{cases}
      u_i(x_i) & \textmd{if }x_i \ge t_i\\
      0 & \textmd{if }x_i < t_i
    \end{cases}\end{equation}
  with
  \begin{equation}
    \label{uti2}
    u_i(x_i) = \begin{cases}
      d & \textmd{if }x_i \ge r_i''\\
      \frac d{r_i''}x_i & \textmd{if }x_i < r_i''
    \end{cases},\end{equation}
  where $d$ is a positive constant. Then for true values $\mathbf
  t_{-i} = \mathbf r_{-i}$, $t_i = r_i'$ and possibly lying bid $s_i =
  r_i''$,
  \begin{align*}
    U_i(p, x, t_i, t_i, \mathbf t_{-i}) & =
    \sum_{\phi:i\in\phi}u_i(t_i, x_{i, \phi}(\mathbf t_{-i},
    t_i))p_\phi(\mathbf t_{-i}, t_i)\\
    & = \sum_{\phi:i\in\phi, \phi\neq\psi}u_i(t_i, x_{i,
      \phi}(\mathbf t_{-i}, t_i))p_\phi(\mathbf t_{-i}, t_i)\\
    & \quad+\:u_i(t_i, x_{i, \psi}(\mathbf t_{-i},
    t_i))p_{\psi}(\mathbf t_{-i}, t_i)\\
    & \le d\sum_{\phi:i\in\phi,
      \phi\neq\psi}p_\phi(\mathbf t_{-i}, t_i)\\
    & \quad+\:\left(\frac d{r_i''}x_{i, \psi}(\mathbf r_{-i}, r_i')\right)p_{\psi}(\mathbf t_{-i}, t_i)\\
    & < d\sum_{\phi:i\in\phi, \phi\neq\psi}p_\phi(\mathbf t_{-i},
    t_i)+dp_\psi(\mathbf t_{-i}, t_i)\\
    & = d\sum_{\phi:i\in\phi}p_\phi(\mathbf t_{-i}, t_i)\\
    & = d = U_i(p, x, t_i, s_i, \mathbf t_{-i})\ .
  \end{align*}
  That is, user $i$ with true value $t_i$ has incentive to lie to bid
  $s_i$.

  For the second case, for any $\psi\in\Psi_i(\mathbf r_{-i}, r_i')$, we have
  $x_{i, \psi}(\mathbf r_{-i}, r_i') \le r_i$. We consider the
  following utility function:
  \begin{equation*}
    u_i(t_i, x_i) =
    \begin{cases}
      d & \textmd{if } x_i \ge t_i\\
      0 & \textmd{if } x_i < t_i
    \end{cases},
  \end{equation*}
  where $d$ is a positive constant. Let the true values $\mathbf
  t_{-i} = \mathbf r_{-i}$, $t_i = r_i$, and the possible lying bid
  $s_i = r_i'$. We have
  \begin{align*}
    U_i(p, x, t_i, t_i, \mathbf t_{-i}) & =
    d\sum_{\phi:i\in\phi}p_\phi(\mathbf t)\\
    & < d = d\sum_{\phi:i\in\phi}p_\phi(\mathbf t_{-i}, s_i) = U_i(p,
    x, t_i, s_i, \mathbf t_{-i})\ .
  \end{align*}
  That is, user $i$ with true value $t_i$ has incentive to lie to bid
  $s_i$. This contradiction completes the proof.
\end{IEEEproof}
\subsection{Proof of Theorem~\ref{lwbthm}}
\label{pflwbthm}
\begin{IEEEproof}
  Here we assume that a user always wins by the bid $s_i = 0$, that
  is, for any $i\in\mathcal N$ and any $\mathbf t_{-i}\in T_{-i}$,
  \begin{equation*}
    \sum_{\phi:i\in\phi}p_\phi(0, \mathbf t_{-i}) = 1\ .
  \end{equation*}
  Indeed, this is reasonable because by bidding a service rate of 0
  the user is willing to drop all packets, thus the AP should always
  be able to admit this user. Also, $x_{i, \phi}(\mathbf t_{-i}, 0)
  \ge 0$ is true, so IR is always satisfied.
  \begin{enumerate}[i.]
  \item Suppose a deterministic mechanism $\mathcal A = (p, x)$
    satisfies IC. Let the winning set function be $\psi(\cdot)$.

    We construct a function $z(\cdot)$ as follows. For any $\mathbf
    t_{-i}\in T_{-i}$, consider the winning bid set for user $i$
    \[W_i(\mathbf t_{-i}) = \{s_i|i\in\psi(\mathbf t_{-i}, s_i)\}\ .\]
    Then $W_i(\mathbf t_{-i})\neq\emptyset$ since $0\in W_i(\mathbf
    t_{-i})$. We assert that for any $s_i\in W_i(\mathbf t_{-i})$, the
    assigned service rate $x_i(\mathbf t_{-i}, s_i)$ is the same (recall that the
    assigned service rate for deterministic mechanisms can be denoted by single
    subscript, as in (\ref{singlesub})). Indeed, if there exist $s_i,
    s_i'\in W_i(\mathbf t_{-i})$ with $x_i(\mathbf t_{-i}, s_i) \neq
    x_i(\mathbf t_{-i}, s_i')$, we must have $s_i \neq s_i'$. We may
    assume $x_i(\mathbf t_{-i}, s_i) > x_i(\mathbf t_{-i}, s_i')$. Let
    $t_i = s_i'$, then given that other users bid $\mathbf t_{-i}$,
    user $i$ with true value $t_i$ has incentive to lie to bid $s_i$,
    since the assigned service rate would be higher. Thus the following function
    \begin{equation*}
      z(\mathbf t_{-i}) = \{x_i(\mathbf t_{-i}, s_i)|s_i\in
      W_i(\mathbf t_{-i})\}
    \end{equation*}
    is well defined.

    Now we want to show that $z(\cdot)$ is the highest winning bid
    function of $(p, x)$. This is equivalent to showing that
    $W_i(\mathbf t_{-i}) = [0, z(\mathbf t_{-i})]$. For any $s_i \le
    z(\mathbf t_{-i})$, if $s_i\notin W_i(\mathbf t_{-i})$, then the
    user with true value $t_i = s_i$ has incentive to bid any winning
    bid $s_i'\in W_i(\mathbf t_{-i})$ since the assigned service rate $z(\mathbf
    t_{-i}) \ge t_i$, which contradicts IC. For any $s_i > z(\mathbf
    t_{-i})$, if $s_i\in W_i(\mathbf t_{-i})$, then the assigned service rate is
    $x_i(\mathbf t_{-i}, s_i) = z(\mathbf t_{-i}) < s_i$, which
    violates IR. Thus $W_i(\mathbf t_{-i}) = [0, z(\mathbf t_{-i})]$,
    that is, $\mathcal A$ is a highest winning bid mechanism with
    highest winning bid function $z(\cdot)$.
  \item Suppose $\mathcal A$ is a highest winning bid mechanism with
    highest winning bid function $z\colon T_{-i}\to T_i$. Then when
    other users' bids $\mathbf t_{-i}$ are fixed, user $i$ with true value
    $t_i \le z(\mathbf t_{-i})$ has no incentive to lie, because
    bidding $s_i \le z(\mathbf t_{-i})$ results in the same assigned service rate
    and bidding $s_i > z(\mathbf t_{-i})$ kicks user $i$
    out. Meanwhile, user $i$ with true value $t_i > z(\mathbf t_{-i})$
    does not have incentive to lie either, since bidding $s_i \le
    z(\mathbf t_{-i})$ results in a assigned service rate too low to accept and
    bidding $s_i > z(\mathbf t_{-i})$ keeps user $i$ out of
    admittance. Thus $\mathcal A$ satisfies IC.
  \end{enumerate}
\end{IEEEproof}
\subsection{Proof of Lemma~\ref{iwbprop}}
\label{pfiwbprop}
\begin{IEEEproof}
  This can be checked by directly going through the process of
  $\bar{\mathcal A}$. If $t_i < \bar z(\mathbf t_{-i})$, we can see
  that $\bar x(\mathbf t) = \bar z(\mathbf t_{-i}) > t_i$, so user $i$
  wins. If $t_i > \bar z(\mathbf t_{-i})$, we have $\bar x(\mathbf t)
  \le t_i$, so user $i$ loses. If $t_i = \bar z(\mathbf t_{-i})$, user
  $i$ wins with some probability between 0 and 1, which depends on the
  number of users bidding $\bar z(\mathbf t_{-i})$.
\end{IEEEproof}
\subsection{Proof of Lemma~\ref{sp}}
\label{pfsp}
\begin{IEEEproof}
  By \eqref{lwbfunc}, we can see that the highest winning bid function of
  $\bar{\mathcal A}^*$ is actually a function of $\sigma_{m_1(\mathbf
    t_{-i})}(\mathbf t_{-i})$ and its ranking $m_1(\mathbf t_{-i})$,
  that is,
  \begin{equation}
    \label{pfsp:eq1}
    \bar z^*(\mathbf t_{-i}) = g\left(\sigma_{m_1(\mathbf t_{-i})}(\mathbf
      t_{-i}), m_1(\mathbf t_{-i})\right)
  \end{equation}
  so the highest winning bid function is determined once the $m_1(\mathbf t_{-i})$'th
  highest bid and its ranking is determined.

  Suppose $\bar{\mathcal A}^*$ is not single-priced, then there exists
  $\mathbf t \in T$ and $i, j\in\mathcal N$ such that
  \begin{equation}
    \label{pfsp:eq2}
    t_i \le \bar z^*(\mathbf t_{-i}), t_j \le \bar z^*(\mathbf
    t_{-j}), \bar z^*(\mathbf t_{-i}) \neq \bar z^*(\mathbf t_{-j})\ .
  \end{equation}
  Given $\tilde{\mathbf t}\in\prod_{k\neq i, k\neq j}T_k$, let
  \[f(t) = \bar z^*(\tilde{\mathbf t}, t)\ ,\] then we first claim that for
  any $\tilde{\mathbf t}$, there exists some $t_0\in [0, 1]$ such that
  \begin{equation}
    \label{pfsp:eq3}
    \begin{cases}
      f(t) \le t & \textmd{for } t > t_0\\
      f(t) = t_0 & \textmd{for } t \le t_0
    \end{cases}\ .
  \end{equation}
  
  First, consider the case $f(0) = 0$. Then for any $t > 0$, if $f(t)
  > t$, then there exists $k\in\{1, 2, \dots, n-1\}$ such that
  \begin{align*}
    \begin{cases}
      \sigma_k(\tilde{\mathbf t}, t) = \bar z(\tilde{\mathbf t}, t)
      \ge \bar z^*(\tilde{\mathbf t}, t) = f(t)\\
      k\sigma_k(\tilde{\mathbf t}, t) \le 1\\
      (k+1)\sigma_{k+1}(\tilde{\mathbf t}, t) > 1
    \end{cases},
  \end{align*}
  which implies, by changing $t$ to 0, we have
  \begin{align*}
    \begin{cases}
      \sigma_k(\tilde{\mathbf t}, 0) = \sigma_k(\tilde{\mathbf t}, t)\\
      k\sigma_k(\tilde{\mathbf t}, 0) =
      k\sigma_k(\tilde{\mathbf t}, t) \le 1\\
      (k+1)\sigma_{k+1}(\tilde{\mathbf t}, 0) =
      (k+1)\sigma_{k+1}(\tilde{\mathbf t}, t) > 1
    \end{cases},
  \end{align*}
  that is,
  \[m_1(\tilde{\mathbf t}, 0) = m_2(\tilde{\mathbf t},
  t)\quad\textmd{and}\quad \sigma_{m_1(\tilde{\mathbf t},
    0)}(\tilde{\mathbf t}, 0) = \sigma_{m_1(\tilde{\mathbf t},
    t)}(\tilde{\mathbf t}, t)\]
  and then by \eqref{pfsp:eq1} we have
  \[f(t) = f(0) = 0 < t,\]
  which contradicts the assumption that $f(t) > t$. So $f(t) \le t$
  for any $t > 0$. Thus we have $t_0 = 0$ in \eqref{pfsp:eq3}.

  Now we consider the case $f(0) > 0$. Then for any $t > f(0)$, if
  $f(t) > t$, then there exists $k\in\{1, 2, \dots, n-1\}$ such that
  \begin{align*}
    \begin{cases}
      \sigma_k(\tilde{\mathbf t}, t) = \bar z(\tilde{\mathbf t}, t)
      \ge \bar z^*(\tilde{\mathbf t}, t) = f(t)\\
      k\sigma_k(\tilde{\mathbf t}, t) \le 1\\
      (k+1)\sigma_{k+1}(\tilde{\mathbf t}, t) > 1
    \end{cases},
  \end{align*}
  which implies, by changing $t$ to 0, we again have
  \begin{align*}
    \begin{cases}
      \sigma_k(\tilde{\mathbf t}, 0) = \sigma_k(\tilde{\mathbf t}, t)\\
      k\sigma_k(\tilde{\mathbf t}, 0) =
      k\sigma_k(\tilde{\mathbf t}, t) \le 1\\
      (k+1)\sigma_{k+1}(\tilde{\mathbf t}, 0) =
      (k+1)\sigma_{k+1}(\tilde{\mathbf t}, t) > 1
    \end{cases},
  \end{align*}
  that is,
  \[m_1(\tilde{\mathbf t}, 0) = m_2(\tilde{\mathbf t},
  t)\quad\textmd{and}\quad \sigma_{m_1(\tilde{\mathbf t},
    0)}(\tilde{\mathbf t}, 0) = \sigma_{m_1(\tilde{\mathbf t},
    t)}(\tilde{\mathbf t}, t)\]
  and again by \eqref{pfsp:eq1} we have
  \[f(t) = f(0) < t,\]
  which contradicts the assumption that $f(t) > t$. So $f(t) \le t$
  for any $t > f(0)$.

  For $t' \le f(0)$, since $f(0) = \sigma_{m_1(\tilde{\mathbf t},
    0)}(\tilde{\mathbf t}, 0)$, we have
  \[m_1(\tilde{\mathbf t}, t') = m_1(\tilde{\mathbf t},
  0)\quad\textmd{and}\quad \sigma_{m_1(\tilde{\mathbf t},
    t')}(\tilde{\mathbf t}, t') = \sigma_{m_1(\tilde{\mathbf t},
    0)}(\tilde{\mathbf t}, 0).\]
  Then by \eqref{pfsp:eq1}, $f(t') = f(0)$ and thus we set $t_0 = f(0)$
  in \eqref{pfsp:eq3}. Therefore the claim is proved.

  We then show that $\bar{\mathcal A}^*$ is single-priced. For any bid
  vector $\mathbf t$, let $\tilde{\mathbf t}$ be the bids in $\mathbf
  t$ other than $t_i$ and $t_j$, and let $t_0$ be such that
  \eqref{pfsp:eq3} holds for $\tilde{\mathbf t}$, then by \eqref{pfsp:eq2}
  \[t_i \le \bar z^*(\mathbf t_{-i}) = \bar z^*(\tilde{\mathbf t},
  t_j) = f(t_j)\]
  and
  \[t_j \le \bar z^*(\mathbf t_{-j}) = \bar z^*(\tilde{\mathbf t},
  t_i) = f(t_i)\ .\]

  First consider the case if $t_i > t_0$ and $t_j > t_0$. We have by
  \eqref{pfsp:eq3}
  \[t_i \ge f(t_i) \ge t_j \ge f(t_j) \ge t_i\]
  and then $t_i = t_j$ and $\bar z^*(\mathbf t_{-i}) = f(t_j) = f(t_i)
  = \bar z^*(\mathbf t_{-j})$.

  Then consider the case if $t_i > t_0$ and $t_j \le t_0$. Then by
  \eqref{pfsp:eq3},
  \[t_0 = f(t_j) \ge t_i\]
  which contradicts $t_0 < t_i$, so this case is not possible.

  Then consider the last case if $t_i \le t_0$ and $t_j \le t_0$. By
  \eqref{pfsp:eq3},
  \[\bar z^*(\mathbf t_{-i}) = f(t_j) = t_0 = f(t_i) = \bar
  z^*(\mathbf t_{-j}).\]

  Thus we always have $\bar z^*(\tilde{\mathbf t}, t_i) = \bar
  z^*(\tilde{\mathbf t}, t_j)$, which contradicts the previous
  assumption. Therefore, $\bar{\mathcal A}^*$ is single-priced.
\end{IEEEproof}
\subsection{Proof of Lemma~\ref{feasibility2}}
\label{pffea2}
\begin{IEEEproof}
  P comes from determinism. IR comes from the definition of highest
  winning bid mechanisms. IC comes from Theorem~\ref{lwbthm}. So only
  the proof of CC requires some more effort.

  For given $\mathbf t$, we divide the winning users into two parts:
  those whose price is dropped, denoted by $A$ and those whose price
  is not dropped, denoted by $B$. Then
  \[A = \{i\in\bar\psi^*(\mathbf t)|\bar z^*(\mathbf t_{-i}) = \bar
  z(\mathbf t_{-i})\}\]
  \[B = \{i\in\bar\psi^*(\mathbf t)|\bar z^*(\mathbf t_{-i}) < \bar
  z(\mathbf t_{-i})\}\] where $\bar\psi^*(\cdot)$ is the winning set
  function for $\bar{\mathcal A}^*$. Then $A\cap B = \emptyset$ and
  $A\cup B = \bar\psi^*(\mathbf t)$.

  If $B = \emptyset$, then for any $i\in\bar\psi^*(\mathbf t)$, let $m
  = m_1(\mathbf t_{-i})$ and we have
  \[(m+1)\sigma_{m+1}(\mathbf t_{-i}) > 1\]
  \[(m+1)\sigma_m(\mathbf t_{-i}) \le 1\]
  then
  \[\sigma_{m+1}(\mathbf t_{-i}) > \sigma_m(\mathbf t_{-i}) = \bar
  z(\mathbf t_{-i}) = \bar z^*(\mathbf t_{-i})\]
  and
  \[|\bar\psi^*(\mathbf t)| \le m+1\ .\]
  Then we have
  \begin{align}
    \label{pffeas2:eq1}
    \sum_{j\in\bar\psi^*(\mathbf t)}\bar x_j^*(\mathbf t) & =
    |\bar\psi^*(\mathbf t)|\cdot\bar z^*(\mathbf t_{-i})\\
    & \le (m+1)\sigma_m(\mathbf t_{-i})\nonumber\\
    & \le 1\nonumber
  \end{align}
  where \eqref{pffeas2:eq1} follows because $\bar{\mathcal A}^*$ is
  single-priced and the other two inequalities come from the analysis
  we have above.

  If $B \neq \emptyset$, then there exists $i\in\bar\psi^*(\mathbf t)$
  such that
  \[t_i \le \bar z^*(\mathbf t_{-i}) < \bar z(\mathbf t_{-i}) =
  \sigma_m(\mathbf t_{-i})\]
  where $m = m_1(\mathbf t_{-i})$. By the definition of $m_1(\cdot)$
  we have
  \[m\sigma_m(\mathbf t_{-i}) \le 1\ .\] By the single price property
  of $\bar{\mathcal A}^*$ we have
  \begin{align}
    \label{pffeas2:eq2}
    |\bar\psi^*(\mathbf t)| & \le |\{j:t_j \le \bar z^*(\mathbf
    t_{-i})\}|\\
    & \le |\{j:t_j < \bar z(\mathbf t_{-i})\}|\nonumber\\
    & \le m\nonumber
  \end{align}
  where \eqref{pffeas2:eq2} comes from the single price property and
  the other two inequalities are due to the analysis above. Again by
  the single price property we have
  \begin{align*}
    \sum_{j\in\bar\psi^*(\mathbf t)}\bar x_j^*(\mathbf t) & =
    |\bar\psi^*(\mathbf t)|\cdot\bar z^*(\mathbf t_{-i})\\
    & \le m\sigma_m(\mathbf t_{-i})\\
    & \le 1\ .
  \end{align*}
  Thus $\bar{\mathcal A}^*$ is feasible.

  From \eqref{lwbfunc} we have
  \[1 \ge \bar z(\mathbf t_{-i}) \ge \bar z^*(\mathbf t_{-i}) \ge
  \sigma_m(\mathbf t_{-i})-\left(\sigma_m(\mathbf t_{-i})-\frac
    1{m+1}\right) \ge 0.\] That is, $0 \le \bar z^*(\mathbf t_{-i}) \le
  1$ for any $\mathbf t_{-i}\in T_{-i}$.
\end{IEEEproof}
\subsection{Proof of Lemma~\ref{wc1}}
\label{pfwc1}
\begin{IEEEproof}
  For any $\mathbf t$, let $m = |\mathcal F(\mathbf t)|$, then
  \[m\sigma_m(\mathbf t) \le 1\]
  and then
  \[(m-1)\sigma_m(\mathbf t) \le 1\] so $|\bar{\mathcal
    A}(\mathbf t)| \ge m-1 = |\mathcal F(\mathbf t)|-1$.
  Also,
  \[(m+1)\sigma_{m+2}(\mathbf t) \ge (m+1)\sigma_{m+1}(\mathbf t) >
  1\] so $|\bar{\mathcal A}(\mathbf t)| \le |\mathcal F(\mathbf
  t)|$.
\end{IEEEproof}
\subsection{Proof of Lemma~\ref{wc2}}
\label{pfwc2}
\begin{IEEEproof}
  Just consider the case when all $n$ users bid $\frac 1m$. Then
  $\mathcal F$ would admit $m$ out of the $n$ users randomly, while
  $\bar{\mathcal A}^*$ cannot admit any of the $n$ users.
\end{IEEEproof}
\subsection{Proof of Theorem~\ref{bp}}
\label{pfbp}
\begin{IEEEproof}
  Let $\mathbf T$ be the vector of random variables with values taken
  in $T$, and as usual let $\mathbf T_i$ and $\mathbf T_{-i}$ be the
  corresponding components. Note that $T\subset[0, 1]^n$. Then the
  probability that $\bar{\mathcal A}^*$ and $\bar{\mathcal A}$ have
  different results is
  \begin{align*}
    D & = \mathrm{Pr}\left(\bar z^*(\mathbf T_{-i}) < \mathbf T_i <
      \bar z(\mathbf T_{-i})\textmd{ for some }i\right)\\
    & \le \sum_{i = 1}^n \mathrm{Pr}\left(\bar z^*(\mathbf
      T_{-i}) < \mathbf T_i < \bar z(\mathbf T_{-i})\right)\\
    & = \sum_{i = 1}^n\int_{B_i}f(\mathbf t)\mathrm d\mathbf t\\
    & \le \sum_{i = 1}^n\mathcal L(B_i)K
  \end{align*}
  where
  \begin{equation*}
    B_i = \{\mathbf t\in T|\bar z^*(\mathbf t_{-i}) < t_i < \bar
    z(\mathbf t_{-i})\}
  \end{equation*}
  and $\mathcal L(\cdot)$ denotes the Lebesgue measure. Then
  \begin{align*}
    \mathcal L(B_i) & = \int_{B_i}1\mathrm d\mathbf t\\
    & = \int_{T_{-i}}\int_{\bar z^*(\mathbf t_{-i})}^{\bar z(\mathbf
      t_{-i})}1\mathrm dt_i\mathrm d\mathbf t_{-i}\\
    & = \int_{\{\sigma_1(\mathbf t_{-i}) > \frac
      12\}}d_1\left(\sigma_1(\mathbf t_{-i})-\frac 12\right)\mathrm d\mathbf t_{-i}\\
    & \quad+\int_{\{\frac 12 \ge\sigma_2(\mathbf t_{-i}) > \frac
      13\}}d_2\left(\sigma_2(\mathbf t_{-i})-\frac 13\right)\mathrm d\mathbf t_{-i}\\
    & \quad+\dots\\
    & \quad+\int_{\{\frac{1}{n-1} \ge\sigma_2(\mathbf t_{-i}) >
      \frac{1}{n}\}}d_{n-1}\left(\sigma_{n-1}(\mathbf t_{-i})-\frac
      1n\right)\mathrm d\mathbf t_{-i}\\
    & \le \frac 12d_1+\left(\frac 23-\frac
      12\right)d_2+\dots+\left(\frac{n-1}{n}-\frac{n-2}{n-1}\right)d_{n-1}\\
    & \le \max_{1\le j\le n-1}d_j\left(1-\frac 1n\right)\\
    & \le \max_{1\le j\le n-1}d_j\ .
  \end{align*}
  Thus
  \[D \le dnK,\] where $d = \max_{1\le j\le n-1}d_j$.
\end{IEEEproof}
\end{ARXIV}
\end{document}